\documentclass[%
 reprint,
amsmath,amssymb,
aps,
prb,
floatfix,
]{revtex4-2}

\usepackage{graphicx}% Include figure files
\usepackage{dcolumn}% Align table columns on decimal point
\usepackage{bm}% bold math
\usepackage{mystyle}

\begin{document}

\preprint{APS/123-QED}

\title{Magnetism from multiparticle ring exchange in moiré Wigner crystals}
\author{Ilya Esterlis}
\author{Alex Levchenko}
\affiliation{Department of Physics, University of Wisconsin-Madison, Madison, Wisconsin 53706, USA}

\date{\today}

\begin{abstract} 
We investigate the multiparticle ring exchange couplings of the two-dimensional triangular Wigner crystal in external commensurate triangular and honeycomb potentials, using a semiclassical approach valid in the regime where Coulomb interactions dominate over electronic kinetic energy. In this limit, increasing the strength of the potential drives a transition from a ferromagnet to a $120^\circ$ Néel antiferromagnet for both external potential types. In the triangular case, we find that the transition occurs already for a weak potential, whereas in the honeycomb case, it occurs when the potential is nearly two orders of magnitude larger. Our results are relevant to the magnetism of  generalized Wigner crystal phases observed at certain rational fillings of the moiré superlattice in transition-metal dichalcogenide heterobilayers. 
\end{abstract}

\maketitle

Current two-dimensional (2D) materials can be synthesized with an adjustable external periodic potential, generated through moiré patterns in twisted multilayers or lattice-mismatched layers (or by mismatching with a substrate). The geometry, strength, and filling (number of electrons per moiré unit cell) of the moiré potential are experimentally controllable and can be tuned to study a variety of correlated electronic phenomena. In the case of transition-metal dichalcogenide (TMD) heterobilayers, electron crystals -- states of matter in which itinerant electrons spontaneously charge-order -- play a particularly prominent role \cite{regan2020,xu2020,Huang2021,Jin2021,li2021,Li2021a,Li2024,shabani2021}, appearing at certain rational fillings of the moiré superlattice. Such moiré electron crystals -- or ``generalized Wigner crystals" \cite{hubbard1978} -- have also garnered significant theoretical interest \cite{pan2020,padhi2021,Matty2022,Kaushal2022,Duran2023,zhou2024,Yang2024_MI,Yang2024_Honeycomb,kumar2024}.

In this Letter, we consider the magnetism induced by multiparticle exchange processes in these moiré electron crystals. We use a semiclassical WKB (or instanton) method to estimate the exchange couplings associated with various ring exchange processes, from which an effective magnetic Hamiltonian is obtained. We consider both triangular and honeycomb moiré potentials and discuss the expected magnetic ground states based on the multiparticle exchange picture. The results of our WKB calculations for the magnetic phases are in agreement with recent quantum Monte Carlo (QMC) studies of moiré two-dimensional electron gases (2DEGs) \cite{Yang2024_MI,Yang2024_Honeycomb}, suggesting that the semiclassical method is a simple and intuitive approach for investigating the magnetic properties of such systems.  

We start with the Hamiltonian of the 2DEG in the presence of an external moiré potential in the form
    \be
    H = \sum_i \frac{\bfp_i^2}{2m}  + V,
    \label{eq:H}
    \ee
where $m$ is band mass and the sum $i$ is over the $N$ electrons in the system. We decompose the potential energy $V$ of the system as a sum of contributions from the electron-electron interaction and the external potential:
    \be
    V =  \frac 12 \sum_{i\neq j} \frac{e^2}{|\bfr_i - \bfr_j|} + \sum_i V_\text{ext}(\bfr_i).
    \ee

In the absence of the external potential ($V_\text{ext}=0$), the ground state of $H$ is determined by the single dimensionless parameter $r_s = V_C/E_F$, where $E_F$ is the Fermi energy and the Coulomb energy $V_C = e^2/a$, with the interparticle distance $a$ related to the 2D density $n$ according to $a = 1/\sqrt{\pi n}$. At large $r_s$, where the Coulomb energy dominates over the kinetic energy, the 2DEG crystallizes into a triangular lattice Wigner crystal (WC), the transition occurring when $r_s \approx 35$ \cite{Tanatar1989,Attaccalite2002,Drummond2009,azadi2024,smith2024}. As $r_s \to \infty$, residual exchange couplings between the electrons localized on the WC lattice lead to a ferromagnetic ground state, as can be deduced from semiclassical considerations (to be described below) \cite{Thouless1965,roger1983,roger1984,katano2000,Ashizawa2000,chakravarty1999,voelker2001}. Upon approaching the melting transition with decreasing $r_s$, the magnetic interactions become highly frustrated \cite{Bernu2001} and the magnetism is more complex \cite{misguich1998,*misguich1999,Bernu2001,Drummond2009,smith2024}.

We now imagine placing the system in the periodic moiré potential 
    \be
    V_\text{ext}(\bfr) = \sum_{|\bfG| = G_0} v_\bfG e^{i\bfG \cdot \bfr},
    \label{eq:V}
    \ee
where $\{ \bfG \}$ are the moiré reciprocal lattice vectors and $G_0$ denotes the length of shortest non-zero reciprocal lattice vectors. 
We assume the potential is commensurate with the triangular WC lattice and denote the (rational) filling of the moiré unit cell by $\nu$. Below we will consider in detail the cases of a triangular moiré lattice with the same periodicity as the WC lattice ($\nu=1$), as well as a honeycomb lattice where the triangular lattice WC resides on one of the honeycomb sublattices ($\nu = 1/2$). We denote the energy scale associated with the external potential by $V_M = |v_\bfG|$ and it will be convenient below to work with the dimensionless parameter $\epsilon = V_M/V_C$ to measure the effects of the moiré potential. The ground state phase diagram of $H$ is determined by the filling $\nu$ and the two parameters $(r_s, \epsilon)$.

The commensurate potential $V_\text{ext}(\bfr)$ will tend to stabilize the WC over the electron liquid for non-zero $\epsilon$. That is, the critical $r_s$ for crystallization will be reduced from its value at $\epsilon = 0$ \footnote{Depending on the nature of the moiré potential, the WC state may be identified via a metal-insulator (or Mott) transition, as in the case of $\nu=1$ on the triangular lattice, or via the breaking of a discrete symmetry of the external potential, like in the case of $\nu=1/3$ on the triangular lattice or $\nu=1/2$ on the honeycomb lattice.}. The presence of the external potential also influences the exchange couplings and the magnetic ground state. These exchange couplings are the main subject of this paper, which we turn to now.

Magnetic couplings of the WC are generated by various ring-exchanges processes, with the exchange dynamics described by the effective magnetic Hamiltonian \cite{Thouless1965,roger1983,roger1984,katano2000,chakravarty1999,voelker2001}
    \be
    H_\text{eff} = \sum_P (-1)^P J_P \hat P.
    \label{eq:Heff}
    \ee
The sum in Eq.~\eqref{eq:Heff} is taken over all cyclic $n$-particle permutations $P$, with $\hat P$ the associated permutation operator acting on the electron spin degrees of freedom. Here we adopt the convention $J_P > 0$, the sign of each term being determined by the signature of the permutation $(-1)^P$; as shown by Thouless \cite{Thouless1965}, exchange of an odd number of particles leads to a ferromagnetic interaction, while exchange of an even number leads to an antiferromagnetic interaction.

At large $r_s$ the couplings may be estimated by multidimensional WKB (or instanton \cite{Coleman1979}) methods \cite{roger1984,katano2000,voelker2001}, which become exact in the extreme dilute limit $r_s \to \infty$. In this approximation the exchange couplings take the form
    \be
    J_P = \hbar \omega_P e^{-S_P/\hbar},
    \label{eq:JP}
    \ee
where 
    \be
    S_P = \int_{\bfR_0}^{P \bfR_0} \dd R \sqrt{2m [V(\bfR)- E_0]}
    \ee
is the action associated to the classical path that effects the permutation (the ground state energy $E_0$ has been subtracted for convenience). Vector $\bfR=(\bfr_1,\bfr_2,\ldots,\bfr_N)$ is the $2N$-component coordinate vector in the configuration space of the $N$ electrons. The integration is between a reference configuration $\bfR_0$ of the electrons on the WC lattice and the permuted configuration $P\bfR_0$; $\dd R$ is the arc length in the configuration space. The prefactor $\hbar \omega_P$  in Eq.~\eqref{eq:JP} is determined by the Gaussian fluctuations about the classical path. 

For finite $r_s$ the semiclassical approach is only approximate, but nevertheless provides a useful estimate of the exchange couplings \footnote{To avoid confusion:  we assume the moiré potential remains commensurate with the triangular lattice WC as $r_s$ is varied. One can imagine that $r_s$ is changed, for example, by tuning the electron charge $e^2$ or $\hbar$ (as opposed to tuning the density $n$, as is usually assumed).}. In the present paper our focus is on the actions $S_P$, as they determine the qualitative behavior of the exchanges at sufficiently large $r_s$; we have not attempted a systematic calculation of the prefactors appearing in Eq.~\eqref{eq:JP}.

The ring exchange couplings for the WC without external potential ($\epsilon = 0$) have been calculated by a number of authors \cite{roger1983,chakravarty1999,voelker2001,kim2024}. The main conclusion is that the actions $S_P$ for permutations involving up to $n=6$ electrons are of comparable magnitude, with the 3-particle exchange action $S_3$ being the smallest. The 3-particle exchange thus dominates as $r_s\to\infty$, leading to a ferromagnetic ground state (according to the Thouless rule). On the other hand, the fact that the different actions have similar magnitudes  implies that the magnetic interactions at finite $r_s$ become highly frustrated, leading to a complex magnetic phase diagram for $r_s$ near the melting point \cite{Bernu2001}. 

In the presence of $V_\text{ext}$, the exchange couplings become functions of the strength of the potential $\epsilon$. The $r_s$ and $\epsilon$ dependence of the classical action may be written $S_P/\hbar = \sqrt r_s \tilde S_P(\epsilon)$, where the dimensionless function $\tilde S_P(\epsilon)$ is explicitly
    \begin{align}
    \tilde S_P(\epsilon) &= \int_{\tilde \bfR_0}^{P \tilde \bfR_0} \dd \tilde R \sqrt{2[\tilde V(\tilde \bfR) - \tilde E_0]},
    \label{eq:SP} \\
    \tilde V(\tilde \bfR) &= \frac 12 \sum_{i\neq j} \frac{1}{|\tilde \bfr_i - \tilde \bfr_j|} + \epsilon \sum_i \sum_{|\bfG| = G_0} e^{i\tilde \bfG \cdot \tilde \bfr_i}.
    \end{align}
Here we introduced dimensionless coordinates $\tilde \bfr_i = \bfr_i/a$, where $a$ is inter-particle distance, and $\tilde E_0 = E_0 / (e^2/a)$. As $\epsilon$ is increased \footnote{In the deep moiré limit corresponding to exceedingly large values of $\epsilon$, a Hubbard model approach the problem is likely to be more appropriate \cite{wu2018}.}, the confining moiré potential inhibits the motion of the electrons away from their equilibrium WC lattice positions and the minimum action process will be that with the shortest exchange path (involving the smallest number of exchanging particles), the 2-particle exchange. When the antiferromagnetic 2-particle exchange is dominant, the ground state is the three-sublattice $120^\circ$ Néel antiferromagnet \cite{bernu1992,*bernu1994,capriotti1999,white2007}. Thus, in the $r_s \to \infty$ limit, we expect a transition from a ferromagnet at small $\epsilon$ to a $120^\circ$ Néel antiferromagnet at larger $\epsilon$.

To investigate when this transition occurs, we have numerically minimized the action Eq.~\eqref{eq:SP} by discretizing the integral into $M$ steps using the trapezoid rule. We allow $N_\text{move}$ electrons to adjust their positions throughout the exchange, with all other electrons fixed at their equilibrium WC positions. We report the minimal actions for a discretization of $M=16$ and $N_\text{move} \approx 50-70$, the latter depending on the specific exchange process under consideration and on the strength of the moiré potential, with a smaller value of $N_\text{move}$ needed for convergence as the potential gets stronger. We have verified that the actions obtained with these parameters are accurate to within 1\%. 
The electrostatic energy in $\tilde V(\tilde \bfR)$ is efficiently computed using the standard Ewald method \cite{bonsall1977}. 

Below, we present our results for $\tilde S_P(\epsilon)$ for the dominant exchange cycles in a triangular moiré potential with $\nu=1$ and honeycomb potential with $\nu=1/2$. Additional data for the exchanges at $\nu=1/3$ filling of the triangular lattice are relegated to Appendix \ref{app:nu_13}, as the results are qualitatively similar to the case of $\nu=1$.  

\begin{figure}%[t!]
    \centering
    \includegraphics[width=\linewidth]{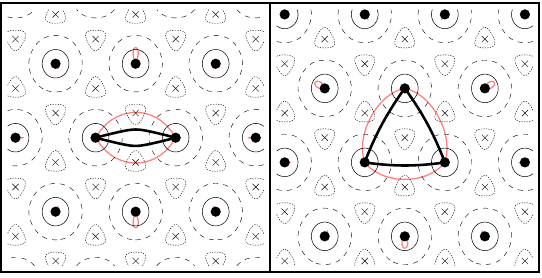}
    \caption{Exchange paths for 2-particle exchange (left) and 3-particle exchange (right) in a triangular moiré potential with one electron per unit cell of the potential ($\nu = 1$ filling). Black circles show positions of the electrons at the moiré potential minima and crosses are maxima of the potential. Thin pink lines and thick black lines indicate paths for $\epsilon = 0$ (no moiré potential) and for $\epsilon \approx 0.26$, respectively.}
    \label{fig:nu1_paths}
\end{figure}

\begin{figure}%[t!]
    \centering
    \includegraphics[width=\linewidth]{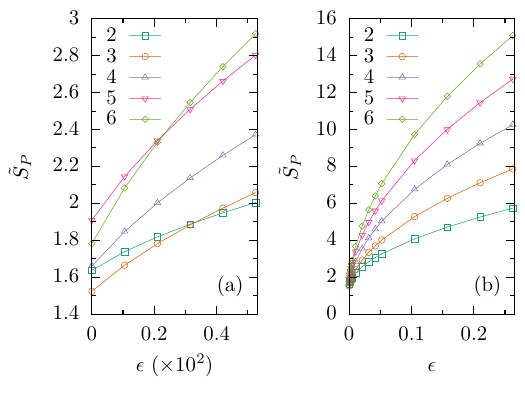}
    \caption{Dimensionless actions $\tilde S_P$ for $n = 2-6$ particle ring-exchanges as a function of the strength of the  triangular moiré potential, $\epsilon$. (a) Weak potential $0 \leq \epsilon \lesssim 0.0053$. (b) Larger range $0 \leq \epsilon \lesssim 0.27$.}
    \label{fig:s_triangular_nu1}
\end{figure}

For the case of the triangular lattice with $\nu=1$,  we have calculated the actions $\tilde S_P$ for the most compact $n=$ 2, 3, 4, 5, and 6 particle ring exchanges as a function of the strength $\epsilon$ of the triangular moiré potential. The 2 and 3-particle exchange paths are shown in Fig.~\ref{fig:nu1_paths}. The presence of the moiré potential ``squeezes" the exchange path away from the maxima of $V_\text{ext}$ and reduces the motion of the non-exchanging electrons away from the equilibrium WC lattice positions. 

The results for the actions $\tilde S_P$ are displayed in Fig.~\ref{fig:s_triangular_nu1}. With increasing $\epsilon$ we find a crossing point between $\tilde S_2$ and $\tilde S_3$ at a relatively small value of the external potential strength $\epsilon = \epsilon_c \approx 0.0033$ ($\tilde S_5$ also becomes smaller than $\tilde S_6$ at a slightly smaller value of $\epsilon \approx 0.0023$). For $\epsilon > \epsilon_c$, $\tilde S_2$ is the smallest action. This implies that, in the large $r_s$ limit, the antiferromagnetic 2-particle exchange is dominant and the ground state will be the $120^\circ$ antiferromagnet for $\epsilon > \epsilon_c$. Fig.~\ref{fig:s_triangular_nu1} also shows that, at larger $\epsilon$,  the magnitude of the action is determined primarily by the length of the exchange path, as anticipated above.

Furthermore, the difference in the magnitudes of the actions shown in Fig.~\ref{fig:s_triangular_nu1} implies a significant reduction in the degree of magnetic frustration as compared to $\epsilon = 0$ (recall the exchange couplings depend exponentially on the actions; see Eq.~\eqref{eq:JP}). The large variations in the magnitudes of the exchange couplings suggests that, even for relatively small values of $\epsilon$, the 120$^\circ$ antiferromagnetic ground state should remain stable all the way to the melting point. For example, when $\epsilon = 0.1$ and $r_s = 30$, we find $J_3/J_2 \sim 10^{-3}$. We have found very similar behavior for $\nu=1/3$ filling of the triangular moiré potential; those results are reported in Appendix \ref{app:nu_13}.

\begin{figure}
    \centering
    \includegraphics[width=\linewidth]{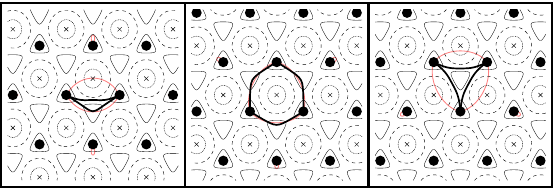}
    \caption{
    Exchange path for 2-particle exchange (left) and two inequivalent 3-particle exchange paths (middle and right) in a honeycomb moiré potential. Paths in the middle and right panels are associated with actions labeled 3 and 3' in Fig.~\ref{fig:s_honeycomb}, respectively. Black circles show positions of the electrons at the moiré potential minima and crosses are maxima of the potential. Thin pink lines and thick black lines indicate paths for $\epsilon = 0$ (no moiré potential) and for $\epsilon \approx 0.53$, respectively.}
    \label{fig:honeycomb_paths}
\end{figure}

\begin{figure}
    \centering
    \includegraphics[width=\linewidth]{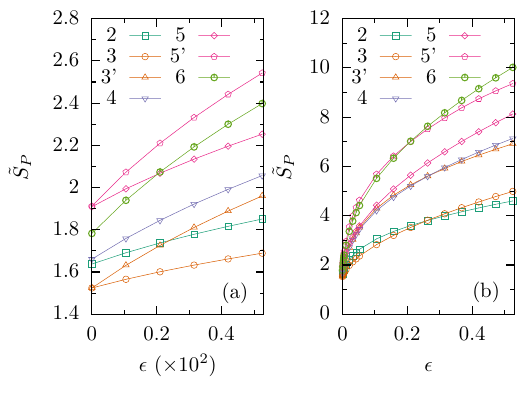}
    \caption{Dimensionless actions $\tilde S_P$ for $n = 2-6$ particle ring-exchanges as a function of the strength of the  honeycomb moiré potential, $\epsilon$. In the honeycomb potential there are two inequivalent 3-particle and 5-particle exchange processes, denoted 3/3' and 5/5'.  (a) Weak potential $0 \leq \epsilon \lesssim 0.0053$. (b) Larger range $0 \leq \epsilon \lesssim 0.53$.}
    \label{fig:s_honeycomb}
\end{figure}

We now turn to the exchange paths and associated actions $\tilde S_P$ in the case of a commensurate honeycomb moiré potential at $\nu=1/2$. The 2 and 3-particle exchange paths are displayed in Fig.~\ref{fig:honeycomb_paths}. In contrast to the triangular moiré potential, there are two inequivalent exchange paths for the 3 and 5 particle exchanges. We denote the additional paths 3' (see Fig.~\ref{fig:honeycomb_paths}) and 5'. 
While the 2-particle exchange path is significantly distorted by the external potential, the 3-particle exchange is only slightly modified because the electrons are able to move near the minima of the moiré potential. This translates into an evolution of the actions that is dramatically different from that observed in the triangular case, as can be seen in Fig.~\ref{fig:s_honeycomb}. The 3-particle action $\tilde S_3$ remains the smallest out to much large value of the moiré potential, with $\tilde S_3$ and $\tilde S_2$ eventually crossing at $\epsilon = \epsilon_c \approx 0.26 $ -- which is nearly two orders of magnitude larger than in the triangular case -- signaling a ferromagnetic to antiferromagnetic transition (at large $r_s$).

In Fig.~\ref{fig:s_honeycomb} we also see that the honeycomb moiré dramatically reduces the importance of the exchange processes involving more than three particles. Thus, even at finite $r_s$, we expect that the competing magnetic phases will primarily be the ferromagnet and 120$^\circ$ antiferromagnet. A precise determination of the corresponding phase boundary requires a more accurate determination of the exchange couplings, beyond the semiclassical approximation \cite{Bernu2001}. 
 
We can compare our results with recent QMC studies on the same model \cite{Yang2024_MI,Yang2024_Honeycomb}. Those calculations were limited to the range $r_s \lesssim 10$, and thus a direct comparison requires significant extrapolation of our semiclassical results. 
Nevertheless, our findings offer a simple rationalization for the QMC results, where it was shown that the triangular moiré stabilizes the 120$^\circ$ antiferromagnetic phase \cite{Yang2024_MI} while the honeycomb moiré stabilizes the ferromagnet \cite{Yang2024_Honeycomb}. Indeed, QMC calculations also suggest that, in the honeycomb case, the ferromagnet remains stable out to relatively large values of the moiré potential \cite{Yang_unpub}, in accord with what we have observed. 

Our study is by no means exhaustive, as we have considered only a limited sampling of moiré potentials and corresponding fillings $\nu$. Experimentally, crystal phases have been observed at a variety of fillings, and the methods we have utilized are immediately applicable to these more general moiré WCs. At generic rational fillings, the triangular WC is not accommodated by the moiré potential, leading to a richer set of possible charge ordered states. For example, at $\nu=1/2$ and $\nu=2/3$ filling of the triangular moiré potential, stripe  \cite{Jin2021,li2021} and honeycomb \cite{li2021} lattices have been observed, respectively. Given the relative simplicity of the approach, an extension of the present calculations to this broader class of moiré WCs certainly seems warranted.

\begin{figure}
    \centering
    \includegraphics[width=\linewidth]{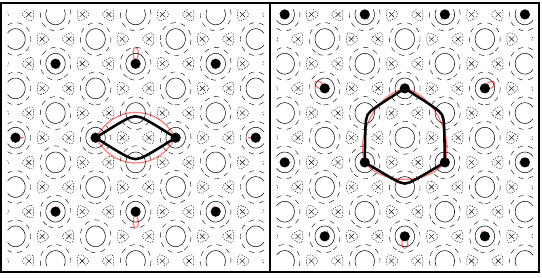}
    \caption{Exchange paths for 2-particle exchange (left) and 3-particle exchange (right) in a triangular moiré potential with $\nu = 1/3$ filling. Black circles show positions of the electrons at the moiré potential minima and crosses are maxima of the potential. Thin pink lines and thick black lines indicate paths for $\epsilon = 0$ (no moiré potential) and for $\epsilon \approx 0.26$, respectively.
    }
    \label{fig:nu13_paths}
\end{figure}

\begin{figure}
    \centering
    \includegraphics[width=\linewidth]{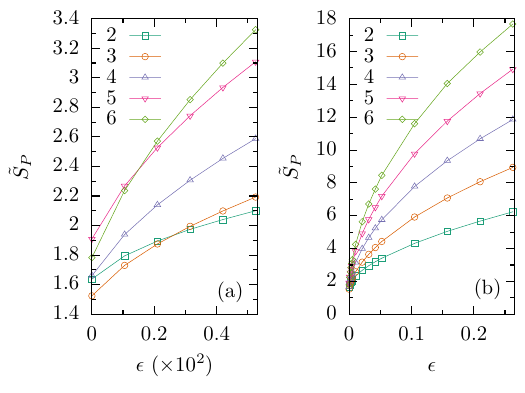}
    \caption{Dimensionless actions $\tilde S_P$ for $n = 2-6$ particle ring-exchanges as a function of the strength of the  triangular moiré potential, $\epsilon$, for $\nu=1/3$ filling. (a) Weak potential $0 \leq \epsilon \lesssim 0.0053$. (b) Larger range $0 \leq \epsilon \lesssim 0.27$.}
    \label{fig:s_triangular_nu13}
\end{figure}

Various generalizations that are relevant to real experimental devices are also easily incorporated in the semiclassical approach. These include screening of the electron-electron interaction by gate electrodes \cite{Valenti2025}, as well as the inclusion of  quenched disorder \cite{voelker2001}. 

A particularly interesting aspect of the moiré WC problem is the effect of slight doping away from commensurate filling, where it is energetically favorable to create a small concentration of interstitials or vacancies. The ``kinetic magnetism" resulting from the dynamics of such defects has recently been predicted to occur with much larger energy scales  than those associated with exchange in both the ordinary WC \cite{kim2022,kim2024,kim2025} and triangular lattice Hubbard models \cite{Morera2023,morera2024}. The interplay between ring exchange processes and kinetic magnetism in such lightly doped moiré WCs is a theoretically interesting and experimentally important question \cite{Tang2020,ciorciaro2023,Tao2024}, and one which -- at least to a first approximation -- is readily addressed with semiclassical methods \cite{kim2022,kim2024}. 

We thank K. S. Kim, Z. Zhuang, and D. Zverevich for valuable discussions, T. Senthil for insights during the FTPI March Meeting Workshop that partially motivated this work, and Y. Yang for sharing his unpublished QMC data with us.
This research was supported by the National Science Foundation (NSF) through the University of Wisconsin Materials Research Science and Engineering Center Grant No. DMR-2309000 (I. E.) and NSF Grant No. DMR-2203411 (A. L.). Support for this research was also provided by the Office of the Vice Chancellor for Research and Graduate Education at the University of Wisconsin–Madison with funding from the Wisconsin Alumni Research Foundation, as well as from the University of Wisconsin – Madison (I. E.) and H. I. Romnes Faculty Fellowship (A. L.).

\appendix

\section{Triangular lattice $\nu = 1/3$}
\label{app:nu_13}

In this Appendix, we report our results for $\nu=1/3$ filling of the triangular moiré lattice. For this particular case we have found a finer discretization of the action integral \eqref{eq:SP} is necessary for convergence at larger values of $\epsilon$. In those cases we use a discretization of $M=32$ steps.

The 2 and 3-particle exchange paths are shown in Fig.~\ref{fig:nu13_paths}, and the actions for $n=2-6$ particle exchanges are shown in Fig.~\ref{fig:s_triangular_nu13}. The evolution is similar to the case $\nu=1$ (see Fig.~\ref{fig:s_triangular_nu1}), with a crossing point between $\tilde S_3$ and $\tilde S_2$ occurring at the relatively small value $\epsilon \approx 0.0026$, signaling the ferromagnet to 120$^\circ$ antiferromagnet transition at large $r_s$. The crossing point between $\tilde S_5$ and $\tilde S_6$ occurs for $\epsilon \approx 0.0015$. Also similar to $\nu=1$, the moiré potential significantly reduces the degree of magnetic frustration, and the antiferromagnetic state is likely to remain stable down to the melting point for even a relatively weak potential. We note that our $\nu=1/3$ results are also consistent with a recent DMRG study \cite{zhou2024}.

\bibliography{moire_wc}

\begin{thebibliography}{53}%
\makeatletter
\providecommand \@ifxundefined [1]{%
 \@ifx{#1\undefined}
}%
\providecommand \@ifnum [1]{%
 \ifnum #1\expandafter \@firstoftwo
 \else \expandafter \@secondoftwo
 \fi
}%
\providecommand \@ifx [1]{%
 \ifx #1\expandafter \@firstoftwo
 \else \expandafter \@secondoftwo
 \fi
}%
\providecommand \natexlab [1]{#1}%
\providecommand \enquote  [1]{``#1''}%
\providecommand \bibnamefont  [1]{#1}%
\providecommand \bibfnamefont [1]{#1}%
\providecommand \citenamefont [1]{#1}%
\providecommand \href@noop [0]{\@secondoftwo}%
\providecommand \href [0]{\begingroup \@sanitize@url \@href}%
\providecommand \@href[1]{\@@startlink{#1}\@@href}%
\providecommand \@@href[1]{\endgroup#1\@@endlink}%
\providecommand \@sanitize@url [0]{\catcode `\\12\catcode `\$12\catcode
  `\&12\catcode `\#12\catcode `\^12\catcode `\_12\catcode `\%12\relax}%
\providecommand \@@startlink[1]{}%
\providecommand \@@endlink[0]{}%
\providecommand \url  [0]{\begingroup\@sanitize@url \@url }%
\providecommand \@url [1]{\endgroup\@href {#1}{\urlprefix }}%
\providecommand \urlprefix  [0]{URL }%
\providecommand \Eprint [0]{\href }%
\providecommand \doibase [0]{https://doi.org/}%
\providecommand \selectlanguage [0]{\@gobble}%
\providecommand \bibinfo  [0]{\@secondoftwo}%
\providecommand \bibfield  [0]{\@secondoftwo}%
\providecommand \translation [1]{[#1]}%
\providecommand \BibitemOpen [0]{}%
\providecommand \bibitemStop [0]{}%
\providecommand \bibitemNoStop [0]{.\EOS\space}%
\providecommand \EOS [0]{\spacefactor3000\relax}%
\providecommand \BibitemShut  [1]{\csname bibitem#1\endcsname}%
\let\auto@bib@innerbib\@empty
%</preamble>
\bibitem [{\citenamefont {Regan}\ \emph {et~al.}(2020)\citenamefont {Regan},
  \citenamefont {Wang}, \citenamefont {Jin}, \citenamefont {{Bakti Utama}},
  \citenamefont {Gao}, \citenamefont {Wei}, \citenamefont {Zhao}, \citenamefont
  {Zhao}, \citenamefont {Zhang}, \citenamefont {Yumigeta}, \citenamefont
  {Blei}, \citenamefont {Carlstr{\"{o}}m}, \citenamefont {Watanabe},
  \citenamefont {Taniguchi}, \citenamefont {Tongay}, \citenamefont {Crommie},
  \citenamefont {Zettl},\ and\ \citenamefont {Wang}}]{regan2020}%
  \BibitemOpen
  \bibfield  {author} {\bibinfo {author} {\bibfnamefont {E.~C.}\ \bibnamefont
  {Regan}}, \bibinfo {author} {\bibfnamefont {D.}~\bibnamefont {Wang}},
  \bibinfo {author} {\bibfnamefont {C.}~\bibnamefont {Jin}}, \bibinfo {author}
  {\bibfnamefont {M.~I.}\ \bibnamefont {{Bakti Utama}}}, \bibinfo {author}
  {\bibfnamefont {B.}~\bibnamefont {Gao}}, \bibinfo {author} {\bibfnamefont
  {X.}~\bibnamefont {Wei}}, \bibinfo {author} {\bibfnamefont {S.}~\bibnamefont
  {Zhao}}, \bibinfo {author} {\bibfnamefont {W.}~\bibnamefont {Zhao}}, \bibinfo
  {author} {\bibfnamefont {Z.}~\bibnamefont {Zhang}}, \bibinfo {author}
  {\bibfnamefont {K.}~\bibnamefont {Yumigeta}}, \bibinfo {author}
  {\bibfnamefont {M.}~\bibnamefont {Blei}}, \bibinfo {author} {\bibfnamefont
  {J.~D.}\ \bibnamefont {Carlstr{\"{o}}m}}, \bibinfo {author} {\bibfnamefont
  {K.}~\bibnamefont {Watanabe}}, \bibinfo {author} {\bibfnamefont
  {T.}~\bibnamefont {Taniguchi}}, \bibinfo {author} {\bibfnamefont
  {S.}~\bibnamefont {Tongay}}, \bibinfo {author} {\bibfnamefont
  {M.}~\bibnamefont {Crommie}}, \bibinfo {author} {\bibfnamefont
  {A.}~\bibnamefont {Zettl}},\ and\ \bibinfo {author} {\bibfnamefont
  {F.}~\bibnamefont {Wang}},\ }\bibfield  {title} {\bibinfo {title} {{Mott and
  generalized Wigner crystal states in WSe$_2$/WS$_2$ moir{\'{e}}
  superlattices}},\ }\href {https://doi.org/10.1038/s41586-020-2092-4}
  {\bibfield  {journal} {\bibinfo  {journal} {Nature}\ }\textbf {\bibinfo
  {volume} {579}},\ \bibinfo {pages} {359} (\bibinfo {year}
  {2020})}\BibitemShut {NoStop}%
\bibitem [{\citenamefont {Xu}\ \emph {et~al.}(2020)\citenamefont {Xu},
  \citenamefont {Liu}, \citenamefont {Rhodes}, \citenamefont {Watanabe},
  \citenamefont {Taniguchi}, \citenamefont {Hone}, \citenamefont {Elser},
  \citenamefont {Mak},\ and\ \citenamefont {Shan}}]{xu2020}%
  \BibitemOpen
  \bibfield  {author} {\bibinfo {author} {\bibfnamefont {Y.}~\bibnamefont
  {Xu}}, \bibinfo {author} {\bibfnamefont {S.}~\bibnamefont {Liu}}, \bibinfo
  {author} {\bibfnamefont {D.~A.}\ \bibnamefont {Rhodes}}, \bibinfo {author}
  {\bibfnamefont {K.}~\bibnamefont {Watanabe}}, \bibinfo {author}
  {\bibfnamefont {T.}~\bibnamefont {Taniguchi}}, \bibinfo {author}
  {\bibfnamefont {J.}~\bibnamefont {Hone}}, \bibinfo {author} {\bibfnamefont
  {V.}~\bibnamefont {Elser}}, \bibinfo {author} {\bibfnamefont {K.~F.}\
  \bibnamefont {Mak}},\ and\ \bibinfo {author} {\bibfnamefont {J.}~\bibnamefont
  {Shan}},\ }\bibfield  {title} {\bibinfo {title} {{Correlated insulating
  states at fractional fillings of moir{\'{e}} superlattices}},\ }\href
  {https://doi.org/10.1038/s41586-020-2868-6} {\bibfield  {journal} {\bibinfo
  {journal} {Nature}\ }\textbf {\bibinfo {volume} {587}},\ \bibinfo {pages}
  {214} (\bibinfo {year} {2020})}\BibitemShut {NoStop}%
\bibitem [{\citenamefont {Huang}\ \emph {et~al.}(2021)\citenamefont {Huang},
  \citenamefont {Wang}, \citenamefont {Miao}, \citenamefont {Wang},
  \citenamefont {Li}, \citenamefont {Lian}, \citenamefont {Taniguchi},
  \citenamefont {Watanabe}, \citenamefont {Okamoto}, \citenamefont {Xiao},
  \citenamefont {Shi},\ and\ \citenamefont {Cui}}]{Huang2021}%
  \BibitemOpen
  \bibfield  {author} {\bibinfo {author} {\bibfnamefont {X.}~\bibnamefont
  {Huang}}, \bibinfo {author} {\bibfnamefont {T.}~\bibnamefont {Wang}},
  \bibinfo {author} {\bibfnamefont {S.}~\bibnamefont {Miao}}, \bibinfo {author}
  {\bibfnamefont {C.}~\bibnamefont {Wang}}, \bibinfo {author} {\bibfnamefont
  {Z.}~\bibnamefont {Li}}, \bibinfo {author} {\bibfnamefont {Z.}~\bibnamefont
  {Lian}}, \bibinfo {author} {\bibfnamefont {T.}~\bibnamefont {Taniguchi}},
  \bibinfo {author} {\bibfnamefont {K.}~\bibnamefont {Watanabe}}, \bibinfo
  {author} {\bibfnamefont {S.}~\bibnamefont {Okamoto}}, \bibinfo {author}
  {\bibfnamefont {D.}~\bibnamefont {Xiao}}, \bibinfo {author} {\bibfnamefont
  {S.-F.}\ \bibnamefont {Shi}},\ and\ \bibinfo {author} {\bibfnamefont {Y.-T.}\
  \bibnamefont {Cui}},\ }\bibfield  {title} {\bibinfo {title} {{Correlated
  insulating states at fractional fillings of the WS$_2$/WSe$_2$ moir{\'{e}}
  lattice}},\ }\href {https://doi.org/10.1038/s41567-021-01171-w} {\bibfield
  {journal} {\bibinfo  {journal} {Nature Physics}\ }\textbf {\bibinfo {volume}
  {17}},\ \bibinfo {pages} {715} (\bibinfo {year} {2021})}\BibitemShut
  {NoStop}%
\bibitem [{\citenamefont {Jin}\ \emph {et~al.}(2021)\citenamefont {Jin},
  \citenamefont {Tao}, \citenamefont {Li}, \citenamefont {Xu}, \citenamefont
  {Tang}, \citenamefont {Zhu}, \citenamefont {Liu}, \citenamefont {Watanabe},
  \citenamefont {Taniguchi}, \citenamefont {Hone}, \citenamefont {Fu},
  \citenamefont {Shan},\ and\ \citenamefont {Mak}}]{Jin2021}%
  \BibitemOpen
  \bibfield  {author} {\bibinfo {author} {\bibfnamefont {C.}~\bibnamefont
  {Jin}}, \bibinfo {author} {\bibfnamefont {Z.}~\bibnamefont {Tao}}, \bibinfo
  {author} {\bibfnamefont {T.}~\bibnamefont {Li}}, \bibinfo {author}
  {\bibfnamefont {Y.}~\bibnamefont {Xu}}, \bibinfo {author} {\bibfnamefont
  {Y.}~\bibnamefont {Tang}}, \bibinfo {author} {\bibfnamefont {J.}~\bibnamefont
  {Zhu}}, \bibinfo {author} {\bibfnamefont {S.}~\bibnamefont {Liu}}, \bibinfo
  {author} {\bibfnamefont {K.}~\bibnamefont {Watanabe}}, \bibinfo {author}
  {\bibfnamefont {T.}~\bibnamefont {Taniguchi}}, \bibinfo {author}
  {\bibfnamefont {J.~C.}\ \bibnamefont {Hone}}, \bibinfo {author}
  {\bibfnamefont {L.}~\bibnamefont {Fu}}, \bibinfo {author} {\bibfnamefont
  {J.}~\bibnamefont {Shan}},\ and\ \bibinfo {author} {\bibfnamefont {K.~F.}\
  \bibnamefont {Mak}},\ }\bibfield  {title} {\bibinfo {title} {{Stripe phases
  in WSe$_2$/WS$_2$ moir{\'{e}} superlattices}},\ }\href
  {https://doi.org/10.1038/s41563-021-00959-8} {\bibfield  {journal} {\bibinfo
  {journal} {Nature Materials}\ }\textbf {\bibinfo {volume} {20}},\ \bibinfo
  {pages} {940} (\bibinfo {year} {2021})}\BibitemShut {NoStop}%
\bibitem [{\citenamefont {Li}\ \emph {et~al.}(2021{\natexlab{a}})\citenamefont
  {Li}, \citenamefont {Li}, \citenamefont {Regan}, \citenamefont {Wang},
  \citenamefont {Zhao}, \citenamefont {Kahn}, \citenamefont {Yumigeta},
  \citenamefont {Blei}, \citenamefont {Taniguchi}, \citenamefont {Watanabe}
  \emph {et~al.}}]{li2021}%
  \BibitemOpen
  \bibfield  {author} {\bibinfo {author} {\bibfnamefont {H.}~\bibnamefont
  {Li}}, \bibinfo {author} {\bibfnamefont {S.}~\bibnamefont {Li}}, \bibinfo
  {author} {\bibfnamefont {E.~C.}\ \bibnamefont {Regan}}, \bibinfo {author}
  {\bibfnamefont {D.}~\bibnamefont {Wang}}, \bibinfo {author} {\bibfnamefont
  {W.}~\bibnamefont {Zhao}}, \bibinfo {author} {\bibfnamefont {S.}~\bibnamefont
  {Kahn}}, \bibinfo {author} {\bibfnamefont {K.}~\bibnamefont {Yumigeta}},
  \bibinfo {author} {\bibfnamefont {M.}~\bibnamefont {Blei}}, \bibinfo {author}
  {\bibfnamefont {T.}~\bibnamefont {Taniguchi}}, \bibinfo {author}
  {\bibfnamefont {K.}~\bibnamefont {Watanabe}}, \emph {et~al.},\ }\bibfield
  {title} {\bibinfo {title} {Imaging two-dimensional generalized {W}igner
  crystals},\ }\href {https://doi.org/10.1038/s41586-021-03874-9} {\bibfield
  {journal} {\bibinfo  {journal} {Nature}\ }\textbf {\bibinfo {volume} {597}},\
  \bibinfo {pages} {650} (\bibinfo {year} {2021}{\natexlab{a}})}\BibitemShut
  {NoStop}%
\bibitem [{\citenamefont {Li}\ \emph {et~al.}(2021{\natexlab{b}})\citenamefont
  {Li}, \citenamefont {Zhu}, \citenamefont {Tang}, \citenamefont {Watanabe},
  \citenamefont {Taniguchi}, \citenamefont {Elser}, \citenamefont {Shan},\ and\
  \citenamefont {Mak}}]{Li2021a}%
  \BibitemOpen
  \bibfield  {author} {\bibinfo {author} {\bibfnamefont {T.}~\bibnamefont
  {Li}}, \bibinfo {author} {\bibfnamefont {J.}~\bibnamefont {Zhu}}, \bibinfo
  {author} {\bibfnamefont {Y.}~\bibnamefont {Tang}}, \bibinfo {author}
  {\bibfnamefont {K.}~\bibnamefont {Watanabe}}, \bibinfo {author}
  {\bibfnamefont {T.}~\bibnamefont {Taniguchi}}, \bibinfo {author}
  {\bibfnamefont {V.}~\bibnamefont {Elser}}, \bibinfo {author} {\bibfnamefont
  {J.}~\bibnamefont {Shan}},\ and\ \bibinfo {author} {\bibfnamefont {K.~F.}\
  \bibnamefont {Mak}},\ }\bibfield  {title} {\bibinfo {title}
  {{Charge-order-enhanced capacitance in semiconductor moir{\'{e}}
  superlattices}},\ }\href {https://doi.org/10.1038/s41565-021-00955-8}
  {\bibfield  {journal} {\bibinfo  {journal} {Nature Nanotechnology}\ }\textbf
  {\bibinfo {volume} {16}},\ \bibinfo {pages} {1068} (\bibinfo {year}
  {2021}{\natexlab{b}})}\BibitemShut {NoStop}%
\bibitem [{\citenamefont {Li}\ \emph {et~al.}(2024)\citenamefont {Li},
  \citenamefont {Xiang}, \citenamefont {Regan}, \citenamefont {Zhao},
  \citenamefont {Sailus}, \citenamefont {Banerjee}, \citenamefont {Taniguchi},
  \citenamefont {Watanabe}, \citenamefont {Tongay}, \citenamefont {Zettl},
  \citenamefont {Crommie},\ and\ \citenamefont {Wang}}]{Li2024}%
  \BibitemOpen
  \bibfield  {author} {\bibinfo {author} {\bibfnamefont {H.}~\bibnamefont
  {Li}}, \bibinfo {author} {\bibfnamefont {Z.}~\bibnamefont {Xiang}}, \bibinfo
  {author} {\bibfnamefont {E.}~\bibnamefont {Regan}}, \bibinfo {author}
  {\bibfnamefont {W.}~\bibnamefont {Zhao}}, \bibinfo {author} {\bibfnamefont
  {R.}~\bibnamefont {Sailus}}, \bibinfo {author} {\bibfnamefont
  {R.}~\bibnamefont {Banerjee}}, \bibinfo {author} {\bibfnamefont
  {T.}~\bibnamefont {Taniguchi}}, \bibinfo {author} {\bibfnamefont
  {K.}~\bibnamefont {Watanabe}}, \bibinfo {author} {\bibfnamefont
  {S.}~\bibnamefont {Tongay}}, \bibinfo {author} {\bibfnamefont
  {A.}~\bibnamefont {Zettl}}, \bibinfo {author} {\bibfnamefont {M.~F.}\
  \bibnamefont {Crommie}},\ and\ \bibinfo {author} {\bibfnamefont
  {F.}~\bibnamefont {Wang}},\ }\bibfield  {title} {\bibinfo {title} {{Mapping
  charge excitations in generalized Wigner crystals}},\ }\href
  {https://doi.org/10.1038/s41565-023-01594-x} {\bibfield  {journal} {\bibinfo
  {journal} {Nature Nanotechnology}\ }\textbf {\bibinfo {volume} {19}},\
  \bibinfo {pages} {618} (\bibinfo {year} {2024})}\BibitemShut {NoStop}%
\bibitem [{\citenamefont {Shabani}\ \emph {et~al.}(2021)\citenamefont
  {Shabani}, \citenamefont {Halbertal}, \citenamefont {Wu}, \citenamefont
  {Chen}, \citenamefont {Liu}, \citenamefont {Hone}, \citenamefont {Yao},
  \citenamefont {Basov}, \citenamefont {Zhu},\ and\ \citenamefont
  {Pasupathy}}]{shabani2021}%
  \BibitemOpen
  \bibfield  {author} {\bibinfo {author} {\bibfnamefont {S.}~\bibnamefont
  {Shabani}}, \bibinfo {author} {\bibfnamefont {D.}~\bibnamefont {Halbertal}},
  \bibinfo {author} {\bibfnamefont {W.}~\bibnamefont {Wu}}, \bibinfo {author}
  {\bibfnamefont {M.}~\bibnamefont {Chen}}, \bibinfo {author} {\bibfnamefont
  {S.}~\bibnamefont {Liu}}, \bibinfo {author} {\bibfnamefont {J.}~\bibnamefont
  {Hone}}, \bibinfo {author} {\bibfnamefont {W.}~\bibnamefont {Yao}}, \bibinfo
  {author} {\bibfnamefont {D.~N.}\ \bibnamefont {Basov}}, \bibinfo {author}
  {\bibfnamefont {X.}~\bibnamefont {Zhu}},\ and\ \bibinfo {author}
  {\bibfnamefont {A.~N.}\ \bibnamefont {Pasupathy}},\ }\bibfield  {title}
  {\bibinfo {title} {{Deep moir{\'{e}} potentials in twisted transition metal
  dichalcogenide bilayers}},\ }\href
  {https://doi.org/10.1038/s41567-021-01174-7} {\bibfield  {journal} {\bibinfo
  {journal} {Nature Physics}\ }\textbf {\bibinfo {volume} {17}},\ \bibinfo
  {pages} {720} (\bibinfo {year} {2021})}\BibitemShut {NoStop}%
\bibitem [{\citenamefont {Hubbard}(1978)}]{hubbard1978}%
  \BibitemOpen
  \bibfield  {author} {\bibinfo {author} {\bibfnamefont {J.}~\bibnamefont
  {Hubbard}},\ }\bibfield  {title} {\bibinfo {title} {Generalized {W}igner
  lattices in one dimension and some applications to tetracyanoquinodimethane
  ({TCNQ}) salts},\ }\href {https://doi.org/10.1103/PhysRevB.17.494} {\bibfield
   {journal} {\bibinfo  {journal} {Phys. Rev. B}\ }\textbf {\bibinfo {volume}
  {17}},\ \bibinfo {pages} {494} (\bibinfo {year} {1978})}\BibitemShut
  {NoStop}%
\bibitem [{\citenamefont {Pan}\ \emph {et~al.}(2020)\citenamefont {Pan},
  \citenamefont {Wu},\ and\ \citenamefont {Das~Sarma}}]{pan2020}%
  \BibitemOpen
  \bibfield  {author} {\bibinfo {author} {\bibfnamefont {H.}~\bibnamefont
  {Pan}}, \bibinfo {author} {\bibfnamefont {F.}~\bibnamefont {Wu}},\ and\
  \bibinfo {author} {\bibfnamefont {S.}~\bibnamefont {Das~Sarma}},\ }\bibfield
  {title} {\bibinfo {title} {Quantum phase diagram of a moir\'e-{H}ubbard
  model},\ }\href {https://doi.org/10.1103/PhysRevB.102.201104} {\bibfield
  {journal} {\bibinfo  {journal} {Phys. Rev. B}\ }\textbf {\bibinfo {volume}
  {102}},\ \bibinfo {pages} {201104} (\bibinfo {year} {2020})}\BibitemShut
  {NoStop}%
\bibitem [{\citenamefont {Padhi}\ \emph {et~al.}(2021)\citenamefont {Padhi},
  \citenamefont {Chitra},\ and\ \citenamefont {Phillips}}]{padhi2021}%
  \BibitemOpen
  \bibfield  {author} {\bibinfo {author} {\bibfnamefont {B.}~\bibnamefont
  {Padhi}}, \bibinfo {author} {\bibfnamefont {R.}~\bibnamefont {Chitra}},\ and\
  \bibinfo {author} {\bibfnamefont {P.~W.}\ \bibnamefont {Phillips}},\
  }\bibfield  {title} {\bibinfo {title} {Generalized {W}igner crystallization
  in moir\'e materials},\ }\href {https://doi.org/10.1103/PhysRevB.103.125146}
  {\bibfield  {journal} {\bibinfo  {journal} {Phys. Rev. B}\ }\textbf {\bibinfo
  {volume} {103}},\ \bibinfo {pages} {125146} (\bibinfo {year}
  {2021})}\BibitemShut {NoStop}%
\bibitem [{\citenamefont {Matty}\ and\ \citenamefont {Kim}(2022)}]{Matty2022}%
  \BibitemOpen
  \bibfield  {author} {\bibinfo {author} {\bibfnamefont {M.}~\bibnamefont
  {Matty}}\ and\ \bibinfo {author} {\bibfnamefont {E.-A.}\ \bibnamefont
  {Kim}},\ }\bibfield  {title} {\bibinfo {title} {{Melting of generalized
  Wigner crystals in transition metal dichalcogenide heterobilayer Moir{\'{e}}
  systems}},\ }\href {https://doi.org/10.1038/s41467-022-34683-x} {\bibfield
  {journal} {\bibinfo  {journal} {Nature Communications}\ }\textbf {\bibinfo
  {volume} {13}},\ \bibinfo {pages} {7098} (\bibinfo {year}
  {2022})}\BibitemShut {NoStop}%
\bibitem [{\citenamefont {Kaushal}\ \emph {et~al.}(2022)\citenamefont
  {Kaushal}, \citenamefont {Morales-Dur{\'{a}}n}, \citenamefont {MacDonald},\
  and\ \citenamefont {Dagotto}}]{Kaushal2022}%
  \BibitemOpen
  \bibfield  {author} {\bibinfo {author} {\bibfnamefont {N.}~\bibnamefont
  {Kaushal}}, \bibinfo {author} {\bibfnamefont {N.}~\bibnamefont
  {Morales-Dur{\'{a}}n}}, \bibinfo {author} {\bibfnamefont {A.~H.}\
  \bibnamefont {MacDonald}},\ and\ \bibinfo {author} {\bibfnamefont
  {E.}~\bibnamefont {Dagotto}},\ }\bibfield  {title} {\bibinfo {title}
  {{Magnetic ground states of honeycomb lattice Wigner crystals}},\ }\href
  {https://doi.org/10.1038/s42005-022-01065-0} {\bibfield  {journal} {\bibinfo
  {journal} {Communications Physics}\ }\textbf {\bibinfo {volume} {5}},\
  \bibinfo {pages} {289} (\bibinfo {year} {2022})}\BibitemShut {NoStop}%
\bibitem [{\citenamefont {Morales-Dur\'an}\ \emph {et~al.}(2023)\citenamefont
  {Morales-Dur\'an}, \citenamefont {Potasz},\ and\ \citenamefont
  {MacDonald}}]{Duran2023}%
  \BibitemOpen
  \bibfield  {author} {\bibinfo {author} {\bibfnamefont {N.}~\bibnamefont
  {Morales-Dur\'an}}, \bibinfo {author} {\bibfnamefont {P.}~\bibnamefont
  {Potasz}},\ and\ \bibinfo {author} {\bibfnamefont {A.~H.}\ \bibnamefont
  {MacDonald}},\ }\bibfield  {title} {\bibinfo {title} {Magnetism and quantum
  melting in moir\'e-material {W}igner crystals},\ }\href
  {https://doi.org/10.1103/PhysRevB.107.235131} {\bibfield  {journal} {\bibinfo
   {journal} {Phys. Rev. B}\ }\textbf {\bibinfo {volume} {107}},\ \bibinfo
  {pages} {235131} (\bibinfo {year} {2023})}\BibitemShut {NoStop}%
\bibitem [{\citenamefont {Zhou}\ \emph {et~al.}(2024)\citenamefont {Zhou},
  \citenamefont {Sheng},\ and\ \citenamefont {Kim}}]{zhou2024}%
  \BibitemOpen
  \bibfield  {author} {\bibinfo {author} {\bibfnamefont {Y.}~\bibnamefont
  {Zhou}}, \bibinfo {author} {\bibfnamefont {D.~N.}\ \bibnamefont {Sheng}},\
  and\ \bibinfo {author} {\bibfnamefont {E.-A.}\ \bibnamefont {Kim}},\
  }\bibfield  {title} {\bibinfo {title} {Quantum melting of generalized
  {W}igner crystals in transition metal dichalcogenide moir\'e systems},\
  }\href {https://doi.org/10.1103/PhysRevLett.133.156501} {\bibfield  {journal}
  {\bibinfo  {journal} {Phys. Rev. Lett.}\ }\textbf {\bibinfo {volume} {133}},\
  \bibinfo {pages} {156501} (\bibinfo {year} {2024})}\BibitemShut {NoStop}%
\bibitem [{\citenamefont {Yang}\ \emph
  {et~al.}(2024{\natexlab{a}})\citenamefont {Yang}, \citenamefont {Morales},\
  and\ \citenamefont {Zhang}}]{Yang2024_MI}%
  \BibitemOpen
  \bibfield  {author} {\bibinfo {author} {\bibfnamefont {Y.}~\bibnamefont
  {Yang}}, \bibinfo {author} {\bibfnamefont {M.~A.}\ \bibnamefont {Morales}},\
  and\ \bibinfo {author} {\bibfnamefont {S.}~\bibnamefont {Zhang}},\ }\bibfield
   {title} {\bibinfo {title} {Metal-insulator transition in a semiconductor
  heterobilayer model},\ }\href
  {https://doi.org/10.1103/PhysRevLett.132.076503} {\bibfield  {journal}
  {\bibinfo  {journal} {Phys. Rev. Lett.}\ }\textbf {\bibinfo {volume} {132}},\
  \bibinfo {pages} {076503} (\bibinfo {year} {2024}{\natexlab{a}})}\BibitemShut
  {NoStop}%
\bibitem [{\citenamefont {Yang}\ \emph
  {et~al.}(2024{\natexlab{b}})\citenamefont {Yang}, \citenamefont {Morales},\
  and\ \citenamefont {Zhang}}]{Yang2024_Honeycomb}%
  \BibitemOpen
  \bibfield  {author} {\bibinfo {author} {\bibfnamefont {Y.}~\bibnamefont
  {Yang}}, \bibinfo {author} {\bibfnamefont {M.~A.}\ \bibnamefont {Morales}},\
  and\ \bibinfo {author} {\bibfnamefont {S.}~\bibnamefont {Zhang}},\ }\bibfield
   {title} {\bibinfo {title} {Ferromagnetic semimetal and charge-density wave
  phases of interacting electrons in a honeycomb moiré potential},\ }\bibfield
   {journal} {\bibinfo  {journal} {Physical Review Letters}\ }\textbf {\bibinfo
  {volume} {133}},\ \href {https://doi.org/10.1103/physrevlett.133.266501}
  {10.1103/physrevlett.133.266501} (\bibinfo {year}
  {2024}{\natexlab{b}})\BibitemShut {NoStop}%
\bibitem [{\citenamefont {Kumar}\ \emph {et~al.}(2024)\citenamefont {Kumar},
  \citenamefont {Lewandowski},\ and\ \citenamefont {Changlani}}]{kumar2024}%
  \BibitemOpen
  \bibfield  {author} {\bibinfo {author} {\bibfnamefont {A.}~\bibnamefont
  {Kumar}}, \bibinfo {author} {\bibfnamefont {C.}~\bibnamefont {Lewandowski}},\
  and\ \bibinfo {author} {\bibfnamefont {H.~J.}\ \bibnamefont {Changlani}},\
  }\href {https://arxiv.org/abs/2409.13814} {\bibinfo {title} {Origin and
  stability of generalized wigner crystallinity in triangular moir\'e systems}}
  (\bibinfo {year} {2024}),\ \Eprint {https://arxiv.org/abs/2409.13814}
  {arXiv:2409.13814 [cond-mat.str-el]} \BibitemShut {NoStop}%
\bibitem [{\citenamefont {Tanatar}\ and\ \citenamefont
  {Ceperley}(1989)}]{Tanatar1989}%
  \BibitemOpen
  \bibfield  {author} {\bibinfo {author} {\bibfnamefont {B.}~\bibnamefont
  {Tanatar}}\ and\ \bibinfo {author} {\bibfnamefont {D.~M.}\ \bibnamefont
  {Ceperley}},\ }\bibfield  {title} {\bibinfo {title} {Ground state of the
  two-dimensional electron gas},\ }\href
  {https://doi.org/10.1103/PhysRevB.39.5005} {\bibfield  {journal} {\bibinfo
  {journal} {Phys. Rev. B}\ }\textbf {\bibinfo {volume} {39}},\ \bibinfo
  {pages} {5005} (\bibinfo {year} {1989})}\BibitemShut {NoStop}%
\bibitem [{\citenamefont {Attaccalite}\ \emph {et~al.}(2002)\citenamefont
  {Attaccalite}, \citenamefont {Moroni}, \citenamefont {Gori-Giorgi},\ and\
  \citenamefont {Bachelet}}]{Attaccalite2002}%
  \BibitemOpen
  \bibfield  {author} {\bibinfo {author} {\bibfnamefont {C.}~\bibnamefont
  {Attaccalite}}, \bibinfo {author} {\bibfnamefont {S.}~\bibnamefont {Moroni}},
  \bibinfo {author} {\bibfnamefont {P.}~\bibnamefont {Gori-Giorgi}},\ and\
  \bibinfo {author} {\bibfnamefont {G.~B.}\ \bibnamefont {Bachelet}},\
  }\bibfield  {title} {\bibinfo {title} {Correlation energy and spin
  polarization in the 2d electron gas},\ }\href
  {https://doi.org/10.1103/PhysRevLett.88.256601} {\bibfield  {journal}
  {\bibinfo  {journal} {Phys. Rev. Lett.}\ }\textbf {\bibinfo {volume} {88}},\
  \bibinfo {pages} {256601} (\bibinfo {year} {2002})}\BibitemShut {NoStop}%
\bibitem [{\citenamefont {Drummond}\ and\ \citenamefont
  {Needs}(2009)}]{Drummond2009}%
  \BibitemOpen
  \bibfield  {author} {\bibinfo {author} {\bibfnamefont {N.~D.}\ \bibnamefont
  {Drummond}}\ and\ \bibinfo {author} {\bibfnamefont {R.~J.}\ \bibnamefont
  {Needs}},\ }\bibfield  {title} {\bibinfo {title} {Phase diagram of the
  low-density two-dimensional homogeneous electron gas},\ }\href
  {https://doi.org/10.1103/PhysRevLett.102.126402} {\bibfield  {journal}
  {\bibinfo  {journal} {Phys. Rev. Lett.}\ }\textbf {\bibinfo {volume} {102}},\
  \bibinfo {pages} {126402} (\bibinfo {year} {2009})}\BibitemShut {NoStop}%
\bibitem [{\citenamefont {Azadi}\ \emph {et~al.}(2024)\citenamefont {Azadi},
  \citenamefont {Drummond},\ and\ \citenamefont {Vinko}}]{azadi2024}%
  \BibitemOpen
  \bibfield  {author} {\bibinfo {author} {\bibfnamefont {S.}~\bibnamefont
  {Azadi}}, \bibinfo {author} {\bibfnamefont {N.~D.}\ \bibnamefont
  {Drummond}},\ and\ \bibinfo {author} {\bibfnamefont {S.~M.}\ \bibnamefont
  {Vinko}},\ }\href {https://arxiv.org/abs/2405.00425} {\bibinfo {title}
  {Quantum {M}onte {C}arlo study of the phase diagram of the two-dimensional
  uniform electron liquid}} (\bibinfo {year} {2024}),\ \Eprint
  {https://arxiv.org/abs/2405.00425} {arXiv:2405.00425 [cond-mat.str-el]}
  \BibitemShut {NoStop}%
\bibitem [{\citenamefont {Smith}\ \emph {et~al.}(2024)\citenamefont {Smith},
  \citenamefont {Chen}, \citenamefont {Levy}, \citenamefont {Yang},
  \citenamefont {Morales},\ and\ \citenamefont {Zhang}}]{smith2024}%
  \BibitemOpen
  \bibfield  {author} {\bibinfo {author} {\bibfnamefont {C.}~\bibnamefont
  {Smith}}, \bibinfo {author} {\bibfnamefont {Y.}~\bibnamefont {Chen}},
  \bibinfo {author} {\bibfnamefont {R.}~\bibnamefont {Levy}}, \bibinfo {author}
  {\bibfnamefont {Y.}~\bibnamefont {Yang}}, \bibinfo {author} {\bibfnamefont
  {M.~A.}\ \bibnamefont {Morales}},\ and\ \bibinfo {author} {\bibfnamefont
  {S.}~\bibnamefont {Zhang}},\ }\href {https://arxiv.org/abs/2405.19397}
  {\bibinfo {title} {Ground state phases of the two-dimension electron gas with
  a unified variational approach}} (\bibinfo {year} {2024}),\ \Eprint
  {https://arxiv.org/abs/2405.19397} {arXiv:2405.19397 [cond-mat.str-el]}
  \BibitemShut {NoStop}%
\bibitem [{\citenamefont {Thouless}(1965)}]{Thouless1965}%
  \BibitemOpen
  \bibfield  {author} {\bibinfo {author} {\bibfnamefont {D.~J.}\ \bibnamefont
  {Thouless}},\ }\bibfield  {title} {\bibinfo {title} {Exchange in solid
  $^3${H}e and the {H}eisenberg {H}amiltonian},\ }\href
  {https://doi.org/10.1088/0370-1328/86/5/301} {\bibfield  {journal} {\bibinfo
  {journal} {Proceedings of the Physical Society}\ }\textbf {\bibinfo {volume}
  {86}},\ \bibinfo {pages} {893} (\bibinfo {year} {1965})}\BibitemShut
  {NoStop}%
\bibitem [{\citenamefont {Roger}\ \emph {et~al.}(1983)\citenamefont {Roger},
  \citenamefont {Hetherington},\ and\ \citenamefont {Delrieu}}]{roger1983}%
  \BibitemOpen
  \bibfield  {author} {\bibinfo {author} {\bibfnamefont {M.}~\bibnamefont
  {Roger}}, \bibinfo {author} {\bibfnamefont {J.~H.}\ \bibnamefont
  {Hetherington}},\ and\ \bibinfo {author} {\bibfnamefont {J.~M.}\ \bibnamefont
  {Delrieu}},\ }\bibfield  {title} {\bibinfo {title} {Magnetism in solid
  $^{3}\mathrm{He}$},\ }\href {https://doi.org/10.1103/RevModPhys.55.1}
  {\bibfield  {journal} {\bibinfo  {journal} {Rev. Mod. Phys.}\ }\textbf
  {\bibinfo {volume} {55}},\ \bibinfo {pages} {1} (\bibinfo {year}
  {1983})}\BibitemShut {NoStop}%
\bibitem [{\citenamefont {Roger}(1984)}]{roger1984}%
  \BibitemOpen
  \bibfield  {author} {\bibinfo {author} {\bibfnamefont {M.}~\bibnamefont
  {Roger}},\ }\bibfield  {title} {\bibinfo {title} {Multiple exchange in
  $^{3}\mathrm{He}$ and in the {W}igner solid},\ }\href
  {https://doi.org/10.1103/PhysRevB.30.6432} {\bibfield  {journal} {\bibinfo
  {journal} {Phys. Rev. B}\ }\textbf {\bibinfo {volume} {30}},\ \bibinfo
  {pages} {6432} (\bibinfo {year} {1984})}\BibitemShut {NoStop}%
\bibitem [{\citenamefont {Katano}\ and\ \citenamefont
  {Hirashima}(2000)}]{katano2000}%
  \BibitemOpen
  \bibfield  {author} {\bibinfo {author} {\bibfnamefont {M.}~\bibnamefont
  {Katano}}\ and\ \bibinfo {author} {\bibfnamefont {D.~S.}\ \bibnamefont
  {Hirashima}},\ }\bibfield  {title} {\bibinfo {title} {Multiple-spin exchange
  in a two-dimensional {W}igner crystal},\ }\href
  {https://doi.org/10.1103/PhysRevB.62.2573} {\bibfield  {journal} {\bibinfo
  {journal} {Phys. Rev. B}\ }\textbf {\bibinfo {volume} {62}},\ \bibinfo
  {pages} {2573} (\bibinfo {year} {2000})}\BibitemShut {NoStop}%
\bibitem [{\citenamefont {Ashizawa}\ and\ \citenamefont
  {Hirashima}(2000)}]{Ashizawa2000}%
  \BibitemOpen
  \bibfield  {author} {\bibinfo {author} {\bibfnamefont {H.}~\bibnamefont
  {Ashizawa}}\ and\ \bibinfo {author} {\bibfnamefont {D.~S.}\ \bibnamefont
  {Hirashima}},\ }\bibfield  {title} {\bibinfo {title} {Wkb calculation of
  multiple spin exchange in monolayer solid ${}^{3}\mathrm{He}$},\ }\href
  {https://doi.org/10.1103/PhysRevB.62.9413} {\bibfield  {journal} {\bibinfo
  {journal} {Phys. Rev. B}\ }\textbf {\bibinfo {volume} {62}},\ \bibinfo
  {pages} {9413} (\bibinfo {year} {2000})}\BibitemShut {NoStop}%
\bibitem [{\citenamefont {Sudip~Chakravarty}\ and\ \citenamefont
  {Voelker}(1999)}]{chakravarty1999}%
  \BibitemOpen
  \bibfield  {author} {\bibinfo {author} {\bibfnamefont {C.~N.}\ \bibnamefont
  {Sudip~Chakravarty}, \bibfnamefont {Steven~Kivelson}}\ and\ \bibinfo {author}
  {\bibfnamefont {K.}~\bibnamefont {Voelker}},\ }\bibfield  {title} {\bibinfo
  {title} {Wigner glass, spin liquids and the metal-insulator transition},\
  }\href {https://doi.org/10.1080/13642819908214845} {\bibfield  {journal}
  {\bibinfo  {journal} {Philosophical Magazine B}\ }\textbf {\bibinfo {volume}
  {79}},\ \bibinfo {pages} {859} (\bibinfo {year} {1999})},\ \Eprint
  {https://arxiv.org/abs/https://doi.org/10.1080/13642819908214845}
  {https://doi.org/10.1080/13642819908214845} \BibitemShut {NoStop}%
\bibitem [{\citenamefont {Voelker}\ and\ \citenamefont
  {Chakravarty}(2001)}]{voelker2001}%
  \BibitemOpen
  \bibfield  {author} {\bibinfo {author} {\bibfnamefont {K.}~\bibnamefont
  {Voelker}}\ and\ \bibinfo {author} {\bibfnamefont {S.}~\bibnamefont
  {Chakravarty}},\ }\bibfield  {title} {\bibinfo {title} {Multiparticle ring
  exchange in the {W}igner glass and its possible relevance to strongly
  interacting two-dimensional electron systems in the presence of disorder},\
  }\href {https://doi.org/10.1103/PhysRevB.64.235125} {\bibfield  {journal}
  {\bibinfo  {journal} {Phys. Rev. B}\ }\textbf {\bibinfo {volume} {64}},\
  \bibinfo {pages} {235125} (\bibinfo {year} {2001})}\BibitemShut {NoStop}%
\bibitem [{\citenamefont {Bernu}\ \emph {et~al.}(2001)\citenamefont {Bernu},
  \citenamefont {C\^andido},\ and\ \citenamefont {Ceperley}}]{Bernu2001}%
  \BibitemOpen
  \bibfield  {author} {\bibinfo {author} {\bibfnamefont {B.}~\bibnamefont
  {Bernu}}, \bibinfo {author} {\bibfnamefont {L.}~\bibnamefont {C\^andido}},\
  and\ \bibinfo {author} {\bibfnamefont {D.~M.}\ \bibnamefont {Ceperley}},\
  }\bibfield  {title} {\bibinfo {title} {Exchange frequencies in the 2d
  {W}igner crystal},\ }\href {https://doi.org/10.1103/PhysRevLett.86.870}
  {\bibfield  {journal} {\bibinfo  {journal} {Phys. Rev. Lett.}\ }\textbf
  {\bibinfo {volume} {86}},\ \bibinfo {pages} {870} (\bibinfo {year}
  {2001})}\BibitemShut {NoStop}%
\bibitem [{\citenamefont {Misguich}\ \emph {et~al.}(1998)\citenamefont
  {Misguich}, \citenamefont {Bernu}, \citenamefont {Lhuillier},\ and\
  \citenamefont {Waldtmann}}]{misguich1998}%
  \BibitemOpen
  \bibfield  {author} {\bibinfo {author} {\bibfnamefont {G.}~\bibnamefont
  {Misguich}}, \bibinfo {author} {\bibfnamefont {B.}~\bibnamefont {Bernu}},
  \bibinfo {author} {\bibfnamefont {C.}~\bibnamefont {Lhuillier}},\ and\
  \bibinfo {author} {\bibfnamefont {C.}~\bibnamefont {Waldtmann}},\ }\bibfield
  {title} {\bibinfo {title} {Spin liquid in the multiple-spin exchange model on
  the triangular lattice: ${}^{3}\mathrm{He}$ on graphite},\ }\href
  {https://doi.org/10.1103/PhysRevLett.81.1098} {\bibfield  {journal} {\bibinfo
   {journal} {Phys. Rev. Lett.}\ }\textbf {\bibinfo {volume} {81}},\ \bibinfo
  {pages} {1098} (\bibinfo {year} {1998})}\BibitemShut {NoStop}%
\bibitem [{\citenamefont {Misguich}\ \emph {et~al.}(1999)\citenamefont
  {Misguich}, \citenamefont {Lhuillier}, \citenamefont {Bernu},\ and\
  \citenamefont {Waldtmann}}]{misguich1999}%
  \BibitemOpen
  \bibfield  {author} {\bibinfo {author} {\bibfnamefont {G.}~\bibnamefont
  {Misguich}}, \bibinfo {author} {\bibfnamefont {C.}~\bibnamefont {Lhuillier}},
  \bibinfo {author} {\bibfnamefont {B.}~\bibnamefont {Bernu}},\ and\ \bibinfo
  {author} {\bibfnamefont {C.}~\bibnamefont {Waldtmann}},\ }\bibfield  {title}
  {\bibinfo {title} {Spin-liquid phase of the multiple-spin exchange
  {H}amiltonian on the triangular lattice},\ }\href
  {https://doi.org/10.1103/PhysRevB.60.1064} {\bibfield  {journal} {\bibinfo
  {journal} {Phys. Rev. B}\ }\textbf {\bibinfo {volume} {60}},\ \bibinfo
  {pages} {1064} (\bibinfo {year} {1999})}\BibitemShut {NoStop}%
\bibitem [{Note1()}]{Note1}%
  \BibitemOpen
  \bibinfo {note} {Depending on the nature of the moiré potential, the WC
  state may be identified via a metal-insulator (or Mott) transition, as in the
  case of $\nu =1$ on the triangular lattice, or via the breaking of a discrete
  symmetry of the external potential, like in the case of $\nu =1/3$ on the
  triangular lattice or $\nu =1/2$ on the honeycomb lattice.}\BibitemShut
  {Stop}%
\bibitem [{\citenamefont {Coleman}(1979)}]{Coleman1979}%
  \BibitemOpen
  \bibfield  {author} {\bibinfo {author} {\bibfnamefont {S.}~\bibnamefont
  {Coleman}},\ }\bibinfo {title} {The uses of instantons},\ in\ \href
  {https://doi.org/10.1007/978-1-4684-0991-8_16} {\emph {\bibinfo {booktitle}
  {The Whys of Subnuclear Physics}}},\ \bibinfo {editor} {edited by\ \bibinfo
  {editor} {\bibfnamefont {A.}~\bibnamefont {Zichichi}}}\ (\bibinfo
  {publisher} {Springer US},\ \bibinfo {address} {Boston, MA},\ \bibinfo {year}
  {1979})\ pp.\ \bibinfo {pages} {805--941}\BibitemShut {NoStop}%
\bibitem [{Note2()}]{Note2}%
  \BibitemOpen
  \bibinfo {note} {To avoid confusion: we assume the moiré potential remains
  commensurate with the triangular lattice WC as $r_s$ is varied. One can
  imagine that $r_s$ is changed, for example, by tuning the electron charge
  $e^2$ or $\hbar $ (as opposed to tuning the density $n$, as is usually
  assumed).}\BibitemShut {Stop}%
\bibitem [{\citenamefont {Kim}\ \emph {et~al.}(2024)\citenamefont {Kim},
  \citenamefont {Esterlis}, \citenamefont {Murthy},\ and\ \citenamefont
  {Kivelson}}]{kim2024}%
  \BibitemOpen
  \bibfield  {author} {\bibinfo {author} {\bibfnamefont {K.-S.}\ \bibnamefont
  {Kim}}, \bibinfo {author} {\bibfnamefont {I.}~\bibnamefont {Esterlis}},
  \bibinfo {author} {\bibfnamefont {C.}~\bibnamefont {Murthy}},\ and\ \bibinfo
  {author} {\bibfnamefont {S.~A.}\ \bibnamefont {Kivelson}},\ }\bibfield
  {title} {\bibinfo {title} {Dynamical defects in a two-dimensional {W}igner
  crystal: Self-doping and kinetic magnetism},\ }\href
  {https://doi.org/10.1103/PhysRevB.109.235130} {\bibfield  {journal} {\bibinfo
   {journal} {Phys. Rev. B}\ }\textbf {\bibinfo {volume} {109}},\ \bibinfo
  {pages} {235130} (\bibinfo {year} {2024})}\BibitemShut {NoStop}%
\bibitem [{Note3()}]{Note3}%
  \BibitemOpen
  \bibinfo {note} {In the deep moiré limit corresponding to exceedingly large
  values of $\epsilon $, a Hubbard model approach the problem is likely to be
  more appropriate \cite {wu2018}.}\BibitemShut {Stop}%
\bibitem [{\citenamefont {Bernu}\ \emph {et~al.}(1992)\citenamefont {Bernu},
  \citenamefont {Lhuillier},\ and\ \citenamefont {Pierre}}]{bernu1992}%
  \BibitemOpen
  \bibfield  {author} {\bibinfo {author} {\bibfnamefont {B.}~\bibnamefont
  {Bernu}}, \bibinfo {author} {\bibfnamefont {C.}~\bibnamefont {Lhuillier}},\
  and\ \bibinfo {author} {\bibfnamefont {L.}~\bibnamefont {Pierre}},\
  }\bibfield  {title} {\bibinfo {title} {Signature of {N}\'eel order in exact
  spectra of quantum antiferromagnets on finite lattices},\ }\href
  {https://doi.org/10.1103/PhysRevLett.69.2590} {\bibfield  {journal} {\bibinfo
   {journal} {Phys. Rev. Lett.}\ }\textbf {\bibinfo {volume} {69}},\ \bibinfo
  {pages} {2590} (\bibinfo {year} {1992})}\BibitemShut {NoStop}%
\bibitem [{\citenamefont {Bernu}\ \emph {et~al.}(1994)\citenamefont {Bernu},
  \citenamefont {Lecheminant}, \citenamefont {Lhuillier},\ and\ \citenamefont
  {Pierre}}]{bernu1994}%
  \BibitemOpen
  \bibfield  {author} {\bibinfo {author} {\bibfnamefont {B.}~\bibnamefont
  {Bernu}}, \bibinfo {author} {\bibfnamefont {P.}~\bibnamefont {Lecheminant}},
  \bibinfo {author} {\bibfnamefont {C.}~\bibnamefont {Lhuillier}},\ and\
  \bibinfo {author} {\bibfnamefont {L.}~\bibnamefont {Pierre}},\ }\bibfield
  {title} {\bibinfo {title} {Exact spectra, spin susceptibilities, and order
  parameter of the quantum {H}eisenberg antiferromagnet on the triangular
  lattice},\ }\href {https://doi.org/10.1103/PhysRevB.50.10048} {\bibfield
  {journal} {\bibinfo  {journal} {Phys. Rev. B}\ }\textbf {\bibinfo {volume}
  {50}},\ \bibinfo {pages} {10048} (\bibinfo {year} {1994})}\BibitemShut
  {NoStop}%
\bibitem [{\citenamefont {Capriotti}\ \emph {et~al.}(1999)\citenamefont
  {Capriotti}, \citenamefont {Trumper},\ and\ \citenamefont
  {Sorella}}]{capriotti1999}%
  \BibitemOpen
  \bibfield  {author} {\bibinfo {author} {\bibfnamefont {L.}~\bibnamefont
  {Capriotti}}, \bibinfo {author} {\bibfnamefont {A.~E.}\ \bibnamefont
  {Trumper}},\ and\ \bibinfo {author} {\bibfnamefont {S.}~\bibnamefont
  {Sorella}},\ }\bibfield  {title} {\bibinfo {title} {Long-range {N}\'eel order
  in the triangular {H}eisenberg model},\ }\href
  {https://doi.org/10.1103/PhysRevLett.82.3899} {\bibfield  {journal} {\bibinfo
   {journal} {Phys. Rev. Lett.}\ }\textbf {\bibinfo {volume} {82}},\ \bibinfo
  {pages} {3899} (\bibinfo {year} {1999})}\BibitemShut {NoStop}%
\bibitem [{\citenamefont {White}\ and\ \citenamefont
  {Chernyshev}(2007)}]{white2007}%
  \BibitemOpen
  \bibfield  {author} {\bibinfo {author} {\bibfnamefont {S.~R.}\ \bibnamefont
  {White}}\ and\ \bibinfo {author} {\bibfnamefont {A.~L.}\ \bibnamefont
  {Chernyshev}},\ }\bibfield  {title} {\bibinfo {title} {Ne\'el order in square
  and triangular lattice {H}eisenberg models},\ }\href
  {https://doi.org/10.1103/PhysRevLett.99.127004} {\bibfield  {journal}
  {\bibinfo  {journal} {Phys. Rev. Lett.}\ }\textbf {\bibinfo {volume} {99}},\
  \bibinfo {pages} {127004} (\bibinfo {year} {2007})}\BibitemShut {NoStop}%
\bibitem [{\citenamefont {Bonsall}\ and\ \citenamefont
  {Maradudin}(1977)}]{bonsall1977}%
  \BibitemOpen
  \bibfield  {author} {\bibinfo {author} {\bibfnamefont {L.}~\bibnamefont
  {Bonsall}}\ and\ \bibinfo {author} {\bibfnamefont {A.~A.}\ \bibnamefont
  {Maradudin}},\ }\bibfield  {title} {\bibinfo {title} {Some static and
  dynamical properties of a two-dimensional {W}igner crystal},\ }\href
  {https://doi.org/10.1103/PhysRevB.15.1959} {\bibfield  {journal} {\bibinfo
  {journal} {Phys. Rev. B}\ }\textbf {\bibinfo {volume} {15}},\ \bibinfo
  {pages} {1959} (\bibinfo {year} {1977})}\BibitemShut {NoStop}%
\bibitem [{Yan()}]{Yang_unpub}%
  \BibitemOpen
  \href@noop {} {}\bibinfo {note} {Y. Yang (private communication)}\BibitemShut
  {NoStop}%
\bibitem [{\citenamefont {Valenti}\ \emph {et~al.}(2025)\citenamefont
  {Valenti}, \citenamefont {Calvera}, \citenamefont {Yang}, \citenamefont
  {Morales}, \citenamefont {Kivelson}, \citenamefont {Esterlis},\ and\
  \citenamefont {Zhang}}]{Valenti2025}%
  \BibitemOpen
  \bibfield  {author} {\bibinfo {author} {\bibfnamefont {A.}~\bibnamefont
  {Valenti}}, \bibinfo {author} {\bibfnamefont {V.}~\bibnamefont {Calvera}},
  \bibinfo {author} {\bibfnamefont {Y.}~\bibnamefont {Yang}}, \bibinfo {author}
  {\bibfnamefont {M.~A.}\ \bibnamefont {Morales}}, \bibinfo {author}
  {\bibfnamefont {S.~A.}\ \bibnamefont {Kivelson}}, \bibinfo {author}
  {\bibfnamefont {I.}~\bibnamefont {Esterlis}},\ and\ \bibinfo {author}
  {\bibfnamefont {S.}~\bibnamefont {Zhang}},\ }\href
  {https://arxiv.org/abs/2501.16430} {\bibinfo {title} {Critical gate distance
  for {W}igner crystallization in the two-dimensional electron gas}} (\bibinfo
  {year} {2025}),\ \Eprint {https://arxiv.org/abs/2501.16430} {arXiv:2501.16430
  [cond-mat.str-el]} \BibitemShut {NoStop}%
\bibitem [{\citenamefont {Kim}\ \emph {et~al.}(2022)\citenamefont {Kim},
  \citenamefont {Murthy}, \citenamefont {Pandey},\ and\ \citenamefont
  {Kivelson}}]{kim2022}%
  \BibitemOpen
  \bibfield  {author} {\bibinfo {author} {\bibfnamefont {K.-S.}\ \bibnamefont
  {Kim}}, \bibinfo {author} {\bibfnamefont {C.}~\bibnamefont {Murthy}},
  \bibinfo {author} {\bibfnamefont {A.}~\bibnamefont {Pandey}},\ and\ \bibinfo
  {author} {\bibfnamefont {S.~A.}\ \bibnamefont {Kivelson}},\ }\bibfield
  {title} {\bibinfo {title} {Interstitial-induced ferromagnetism in a
  two-dimensional {W}igner crystal},\ }\href
  {https://doi.org/10.1103/PhysRevLett.129.227202} {\bibfield  {journal}
  {\bibinfo  {journal} {Phys. Rev. Lett.}\ }\textbf {\bibinfo {volume} {129}},\
  \bibinfo {pages} {227202} (\bibinfo {year} {2022})}\BibitemShut {NoStop}%
\bibitem [{\citenamefont {Kim}\ and\ \citenamefont {Elser}(2025)}]{kim2025}%
  \BibitemOpen
  \bibfield  {author} {\bibinfo {author} {\bibfnamefont {K.-S.}\ \bibnamefont
  {Kim}}\ and\ \bibinfo {author} {\bibfnamefont {V.}~\bibnamefont {Elser}},\
  }\href {https://arxiv.org/abs/2412.03638} {\bibinfo {title} {Itinerant
  ferromagnetism from one-dimensional mobility}} (\bibinfo {year} {2025}),\
  \Eprint {https://arxiv.org/abs/2412.03638} {arXiv:2412.03638
  [cond-mat.str-el]} \BibitemShut {NoStop}%
\bibitem [{\citenamefont {Morera}\ \emph {et~al.}(2023)\citenamefont {Morera},
  \citenamefont {Kan\'asz-Nagy}, \citenamefont {Smolenski}, \citenamefont
  {Ciorciaro}, \citenamefont {Imamoğlu},\ and\ \citenamefont
  {Demler}}]{Morera2023}%
  \BibitemOpen
  \bibfield  {author} {\bibinfo {author} {\bibfnamefont {I.}~\bibnamefont
  {Morera}}, \bibinfo {author} {\bibfnamefont {M.}~\bibnamefont
  {Kan\'asz-Nagy}}, \bibinfo {author} {\bibfnamefont {T.}~\bibnamefont
  {Smolenski}}, \bibinfo {author} {\bibfnamefont {L.}~\bibnamefont
  {Ciorciaro}}, \bibinfo {author} {\bibfnamefont {A.}~\bibnamefont
  {Imamoğlu}},\ and\ \bibinfo {author} {\bibfnamefont {E.}~\bibnamefont
  {Demler}},\ }\bibfield  {title} {\bibinfo {title} {High-temperature kinetic
  magnetism in triangular lattices},\ }\href
  {https://doi.org/10.1103/PhysRevResearch.5.L022048} {\bibfield  {journal}
  {\bibinfo  {journal} {Phys. Rev. Res.}\ }\textbf {\bibinfo {volume} {5}},\
  \bibinfo {pages} {L022048} (\bibinfo {year} {2023})}\BibitemShut {NoStop}%
\bibitem [{\citenamefont {Morera}\ and\ \citenamefont
  {Demler}(2024)}]{morera2024}%
  \BibitemOpen
  \bibfield  {author} {\bibinfo {author} {\bibfnamefont {I.}~\bibnamefont
  {Morera}}\ and\ \bibinfo {author} {\bibfnamefont {E.}~\bibnamefont
  {Demler}},\ }\href {https://arxiv.org/abs/2402.14074} {\bibinfo {title}
  {Itinerant magnetism and magnetic polarons in the triangular lattice hubbard
  model}} (\bibinfo {year} {2024}),\ \Eprint {https://arxiv.org/abs/2402.14074}
  {arXiv:2402.14074 [cond-mat.str-el]} \BibitemShut {NoStop}%
\bibitem [{\citenamefont {Tang}\ \emph {et~al.}(2020)\citenamefont {Tang},
  \citenamefont {LI}, \citenamefont {Li}, \citenamefont {Xu}, \citenamefont
  {Liu}, \citenamefont {Barmak}, \citenamefont {Watanabe}, \citenamefont
  {Taniguchi}, \citenamefont {MacDonald}, \citenamefont {Shan},\ and\
  \citenamefont {Mak}}]{Tang2020}%
  \BibitemOpen
  \bibfield  {author} {\bibinfo {author} {\bibfnamefont {Y.}~\bibnamefont
  {Tang}}, \bibinfo {author} {\bibfnamefont {L.}~\bibnamefont {LI}}, \bibinfo
  {author} {\bibfnamefont {T.}~\bibnamefont {Li}}, \bibinfo {author}
  {\bibfnamefont {Y.}~\bibnamefont {Xu}}, \bibinfo {author} {\bibfnamefont
  {S.}~\bibnamefont {Liu}}, \bibinfo {author} {\bibfnamefont {K.}~\bibnamefont
  {Barmak}}, \bibinfo {author} {\bibfnamefont {K.}~\bibnamefont {Watanabe}},
  \bibinfo {author} {\bibfnamefont {T.}~\bibnamefont {Taniguchi}}, \bibinfo
  {author} {\bibfnamefont {A.~H.}\ \bibnamefont {MacDonald}}, \bibinfo {author}
  {\bibfnamefont {J.}~\bibnamefont {Shan}},\ and\ \bibinfo {author}
  {\bibfnamefont {K.~F.}\ \bibnamefont {Mak}},\ }\bibfield  {title} {\bibinfo
  {title} {{Simulation of Hubbard model physics in WSe$_2$/WS$_2$ moir{\'{e}}
  superlattices}},\ }\href {https://doi.org/10.1038/s41586-020-2085-3}
  {\bibfield  {journal} {\bibinfo  {journal} {Nature}\ }\textbf {\bibinfo
  {volume} {579}},\ \bibinfo {pages} {353} (\bibinfo {year}
  {2020})}\BibitemShut {NoStop}%
\bibitem [{\citenamefont {Ciorciaro}\ \emph {et~al.}(2023)\citenamefont
  {Ciorciaro}, \citenamefont {Smole{\'{n}}ski}, \citenamefont {Morera},
  \citenamefont {Kiper}, \citenamefont {Hiestand}, \citenamefont {Kroner},
  \citenamefont {Zhang}, \citenamefont {Watanabe}, \citenamefont {Taniguchi},
  \citenamefont {Demler},\ and\ \citenamefont {İmamoğlu}}]{ciorciaro2023}%
  \BibitemOpen
  \bibfield  {author} {\bibinfo {author} {\bibfnamefont {L.}~\bibnamefont
  {Ciorciaro}}, \bibinfo {author} {\bibfnamefont {T.}~\bibnamefont
  {Smole{\'{n}}ski}}, \bibinfo {author} {\bibfnamefont {I.}~\bibnamefont
  {Morera}}, \bibinfo {author} {\bibfnamefont {N.}~\bibnamefont {Kiper}},
  \bibinfo {author} {\bibfnamefont {S.}~\bibnamefont {Hiestand}}, \bibinfo
  {author} {\bibfnamefont {M.}~\bibnamefont {Kroner}}, \bibinfo {author}
  {\bibfnamefont {Y.}~\bibnamefont {Zhang}}, \bibinfo {author} {\bibfnamefont
  {K.}~\bibnamefont {Watanabe}}, \bibinfo {author} {\bibfnamefont
  {T.}~\bibnamefont {Taniguchi}}, \bibinfo {author} {\bibfnamefont
  {E.}~\bibnamefont {Demler}},\ and\ \bibinfo {author} {\bibfnamefont
  {A.}~\bibnamefont {İmamoğlu}},\ }\bibfield  {title} {\bibinfo {title}
  {{Kinetic magnetism in triangular moir{\'{e}} materials}},\ }\href
  {https://doi.org/10.1038/s41586-023-06633-0} {\bibfield  {journal} {\bibinfo
  {journal} {Nature}\ }\textbf {\bibinfo {volume} {623}},\ \bibinfo {pages}
  {509} (\bibinfo {year} {2023})}\BibitemShut {NoStop}%
\bibitem [{\citenamefont {Tao}\ \emph {et~al.}(2024)\citenamefont {Tao},
  \citenamefont {Zhao}, \citenamefont {Shen}, \citenamefont {Li}, \citenamefont
  {Kn\"uppel}, \citenamefont {Watanabe}, \citenamefont {Taniguchi},
  \citenamefont {Shan},\ and\ \citenamefont {Mak}}]{Tao2024}%
  \BibitemOpen
  \bibfield  {author} {\bibinfo {author} {\bibfnamefont {Z.}~\bibnamefont
  {Tao}}, \bibinfo {author} {\bibfnamefont {W.}~\bibnamefont {Zhao}}, \bibinfo
  {author} {\bibfnamefont {B.}~\bibnamefont {Shen}}, \bibinfo {author}
  {\bibfnamefont {T.}~\bibnamefont {Li}}, \bibinfo {author} {\bibfnamefont
  {P.}~\bibnamefont {Kn\"uppel}}, \bibinfo {author} {\bibfnamefont
  {K.}~\bibnamefont {Watanabe}}, \bibinfo {author} {\bibfnamefont
  {T.}~\bibnamefont {Taniguchi}}, \bibinfo {author} {\bibfnamefont
  {J.}~\bibnamefont {Shan}},\ and\ \bibinfo {author} {\bibfnamefont {K.~F.}\
  \bibnamefont {Mak}},\ }\bibfield  {title} {\bibinfo {title} {{Observation of
  spin polarons in a frustrated moir{\'{e}} Hubbard system}},\ }\href
  {https://doi.org/10.1038/s41567-024-02434-y} {\bibfield  {journal} {\bibinfo
  {journal} {Nature Physics}\ }\textbf {\bibinfo {volume} {20}},\ \bibinfo
  {pages} {783} (\bibinfo {year} {2024})}\BibitemShut {NoStop}%
\bibitem [{\citenamefont {Wu}\ \emph {et~al.}(2018)\citenamefont {Wu},
  \citenamefont {Lovorn}, \citenamefont {Tutuc},\ and\ \citenamefont
  {MacDonald}}]{wu2018}%
  \BibitemOpen
  \bibfield  {author} {\bibinfo {author} {\bibfnamefont {F.}~\bibnamefont
  {Wu}}, \bibinfo {author} {\bibfnamefont {T.}~\bibnamefont {Lovorn}}, \bibinfo
  {author} {\bibfnamefont {E.}~\bibnamefont {Tutuc}},\ and\ \bibinfo {author}
  {\bibfnamefont {A.~H.}\ \bibnamefont {MacDonald}},\ }\bibfield  {title}
  {\bibinfo {title} {Hubbard model physics in transition metal dichalcogenide
  moir\'e bands},\ }\href {https://doi.org/10.1103/PhysRevLett.121.026402}
  {\bibfield  {journal} {\bibinfo  {journal} {Phys. Rev. Lett.}\ }\textbf
  {\bibinfo {volume} {121}},\ \bibinfo {pages} {026402} (\bibinfo {year}
  {2018})}\BibitemShut {NoStop}%
\end{thebibliography}%

\end{document}